\documentclass[twocolumn]{aastex62}

\usepackage{newtxtext,newtxmath}

\usepackage[T1]{fontenc}
\usepackage{ae,aecompl,ulem,color}

\usepackage[utf8]{inputenc}
\usepackage{graphicx}
\usepackage{amssymb}
\usepackage{amsmath}
\usepackage{xcolor}
\usepackage{natbib}
\usepackage{ulem}

\newcommand{\coeff}[2]{\begin{Bmatrix} #1\\[0.5ex] #2 \end{Bmatrix}}

\shorttitle{Implication of Sub-TeV Emission}
\shortauthors{Derishev \& Piran}

\begin{document}

\title[Implication of Sub-TeV Emission]{The physical conditions of the afterglow  implied by MAGIC's  sub-TeV observations of  GRB 190114C}
\author{Evgeny  Derishev}
\affil{Institute of Applied Physics RAS, 46 Ulyanov st,603950 \\
Nizhny Novgorod, Russia}
\author{Tsvi Piran}
\affiliation{Racah Institute of Physics \\
The Hebrew University of
			Jerusalem, Jerusalem 91904, Israel}

\begin{abstract}
MAGIC's observations of late sub-TeV photons from GRB190114C enable us, for the first time, to determine the details of the emission process in a GRB afterglow  and to pin down the physical parameters, such as the bulk Lorentz factor and the Lorentz factor of the emitting electrons as well as {some of the } microphysical parameters.  We find  that the sub-TeV emission is {Synchrotron-Self Compton (SSC)}  radiation produced at the early afterglow stage. Combining  the  sub-TeV and X-ray observations   we narrow uncertainties in the conditions inside the emitting zone, almost eliminating them for some parameters.  {Seventy}  seconds after the trigger the external shock had Lorentz factor $\simeq 100$ and the electrons producing the observed sub-TeV radiation had a Lorentz factor $\simeq 10^4$,  so that the sub-TeV radiation originates from Comptonization of X-ray photons at the border between Thomson and Klein-Nishina regimes.  The inferred conditions within the emitting zone are  at odds with {theoretical expectations }  unless one assumes moderate (with $\tau \simeq 2$) absorption of sub-TeV photons inside the source. With this correction the conditions are in good agreement with predictions of the pair-balance model, but are also acceptable for generic afterglow model as one of many possibilities.   The different temporal evolution of the IC  peak energy
of  these two models   opens a way to discriminate between them  once late-time detection in the  TeV range become available.
\end{abstract}

\section{Introduction}
\label{sec:introduction}

The bright gamma-ray burst, GRB 190114C, was detected by  {\it Swift}/BAT \citep{Gropp2019}, {\it Fermi}/GBM  \citep{Hamburg2019} and Konus-Wind \citep{Frederiks2019}.  {50}  seconds after the trigger, the MAGIC Cherenkov telescope detected a photon at energy  {above 300~GeV} with  more than 20~$\sigma$ significance \citep{Mirzoyan2019}. 
While GRB $\gamma$-rays of several dozen GeV have been detected in the past  by EGRET \citep{Hurley1994} and  {{\it Fermi}/LAT}  \citep{Abdo2009,Ackermann2014}, this was the first ever  detection of a GRB by a Cherenkov telescope at sub-TeV. With a redshift of $z = 0.4245$  \citep{Selsing2019} the corresponding energy {in the source frame} of the highest  energy photon was $\simeq 0.5 $ TeV. Even though the information at this stage is limited to GCN circulars, we explore here the origin of the sub-TeV emission and show that it can lead to a deep insight on the conditions within the emitting regions.  As we show here these observations enables us to determine the bulk Lorentz factor and the Lorentz factor of the emitting electrons as well as the details of the emission process. We can also  put  limits on the microphysical equipartiton parameters.   
This is the first time that the conditions within the emitting region of a GRB have been determined with such a confidence.

 In previous bursts, the EGRET and {{\it Fermi}/LAT}  GeV emission continued  long  after the prompt emission had faded away and have shown a gradual temporal decay. This has led to the suggestion that the GeV photons arise from the afterglow \citep{Kumar2009,Ghisellini2010,Kumar2010}. As the MAGIC sub-TeV radiation shows a similar behaviour, i.e. it was observed after the bulk of prompt emission had faded away, we explore here the possibility that this  is a part of the early afterglow.

Within the standard afterglow model \citep{Sari1998} the emission is produced via the  synchrotron mechanism in an external shock.  However, 
 emission of sub-TeV photons via synchrotron {mechanism}  is  problematic within  this model \citep{Piran2010}. With any reasonable bulk Lorentz factor of the emitting region, the observed photon energy  violates the burn-off limit \citep[e.g.][]{Guilbert1983,de-Jager1996,Aharonian2000}.  The  most natural emission mechanism is,  therefore, Inverse Compton (IC). We do not consider here other emission mechanisms since Synchrotron-Self-Compton (SSC) is the simplest  option. 

A synchrotron-emitting source must produce high-energy radiation  through upscattering of synchrotron photons by the same electrons. 
Thus, GeV and TeV radiation is expected from GRBs' afterglows, both at early and late stages \citep[e.g.][]{Meszaros1994,Waxman1997,Wei1998,Chiang1999,Panaitescu2000,SariEsin2001,Zhang2001,Guetta2003,Ando2008,Fan2008}. For  typical { early afterglow } parameters at least {a few} per cent of the total power must be transferred to  TeV photons and  this component may even be dominant {\citep{theoreticalTeVlimit}}. Yet multiple attempts to detect GRBs' TeV emission with Cherenkov telescopes resulted only in upper limits  {\citep{NonDetectionHESS1, NonDetectionMAGIC1, NonDetectionWhipple, NonDetectionVERITAS, NonDetectionHESS2, NonDetectionMAGIC2}}. The striking contrast between the theoretical predictions and the actual observations until now remained a puzzle, implying that either the physics of the emitting zone in GRB afterglows is poorly understood (and  no TeV radiation produced for one or another reason) or that TeV radiation was strongly absorbed in all cases of non-detection {\citep[see also ][for a discussion]{Vurm2017}}. {We demonstrate that the}  MAGIC detection  solves the puzzle: GRB afterglows are indeed routinely producing    TeV emission, but it is attenuated by internal absorption, that is usually strong enough to prevent detection.

Consider an IC photon with energy ${ E}_\mathrm{_{IC}}$. The emitting electron must  be energetic enough to upscatter it, namely:
$\Gamma  \gamma_e m_e c^2 \ge { E}_\mathrm{_{IC}}$, where $\Gamma$ is the bulk Lorentz factor, $\gamma_e$ is the Lorentz factor of the accelerated electrons, $m_e$ the electron's rest mass and $c$ the speed of light.  Thus, the sub-TeV observations imply that $\Gamma  \gamma_e \ge 10^6$.
The shock dynamics with reasonable circum-burst densities (see \S \ref{sec:auxiliary_relations})  suggests that the bulk Lorenz factor at the time of the observations  can hardly be above $\sim 100$.  This immediately implies that the typical electron's Lorentz factor satisfies $\gamma_e \ge 10^4$.  

The X-ray  flux observed in GRB 190114C  is  larger but comparable  to the  sub-TeV flux observed at the same time (see \S \ref{sec:observations}). 
Comptonization of seed photons that are more energetic than the X-rays  would be much less efficient  due to Klein-Nishina (KN) effect and the paucity of such photons.  Lower energy seed photons require higher electron Lorentz factors, that  are  ruled out as we show in \S \ref{sec:electron_Lorentz_factor}.
This suggests that the seed photons are X-rays, that we take to be $E_X \sim 10$keV.
In this case $\gamma_e \simeq 10^4 $ satisfies the Thomson IC relation 
${E}_{_\mathrm{IC}, Thomson} \simeq  \gamma_e^2  E_{X}$ and with $\Gamma \simeq 100$ also the KN one ${E}_{_\mathrm{IC, KN}} \simeq \Gamma  \gamma_e m_e c^2$ (see \S \ref{sec:contra-KN_argument} and \ref{sec:electron_Lorentz_factor}), {implying} that the IC process operates on the boundary between the two regimes (Thomson and KN). A careful examination (see \S \ref{sec:contra-KN_argument}) shows that it is in the former and just slightly below the transition region. 

When discussing the implications of our results we consider two afterglow models. Both share the same {shock deceleration dynamics}. 
The first, widely used, ``generic"  model \citep[see e.g.][]{Sari1998, Piran99}  assumes 
that the downstream electrons carry a constant fraction, $\epsilon_e$  of the total downstream energy. This leads to an average Lorentz factor of the electrons, $\gamma_e$ that is  proportional 
to  the bulk Lorentz factor $\Gamma$.

The second is the ``pair balance" model \citep{Derishev2016}, which makes a specific prediction about the average Lorentz factor of radiating electrons along with few other predictions.
This model includes an ``accelerator", which supplies energy to radiating particles, and an ``emitter", which drains energy from the particles and transfers it to synchrotron and IC radiation. The interaction between the two is in the form of two-photon pair production due to internal absorption of high-energy IC photons by low-energy target photons (of synchrotron origin). 
The pair balance model also specifies the energy coming from the magnetic field decay as the source of power for accelerating particles, that leads to the prediction that the Compton $y$ parameter in the emitting zone is a few. 
The need to balance the accelerator's power by {pair loading} 
results in the requirement that the emitting zone is not entirely transparent to  its own IC radiation.
It also drives the Lorentz factor of radiating electrons to a value, which corresponds to the border between Thomson and KN Comptonization regimes of  their own synchrotron radiation.

We begin with a brief summary of the  relevant observations  in \S \ref{sec:observations}. These are the preliminary results as described in  GCNs. Clearly, exact numerical values may change {in the final release of observational data,} but  
this will not undermine the validity of our analysis. Some of the 
conclusions may change, though, if the ratio between the {sub-TeV} and {X-ray luminosities} will be revised significantly. We  outline  in \S \ref{sec:model} the basic emission zone model that we use.  We recapitulate in \S \ref{sec:auxiliary_relations}  essential  results of the blast wave dynamics.  
We estimate in \S \ref{sec:TeV_transparency}  the opacity to {two-photon absorption} of the sub-TeV photons within the emitting region.  
 We investigate in \S \ref{sec:contra-KN_argument} the different IC regimes (Thomson and KN). 
In \S \ref{sec:electron_Lorentz_factor}    
 we constrain the Lorentz factor of the emitting electrons.
In \S \ref{sec:fast_cooling} we derive limits on the cooling rate of the sub-TeV-producing electrons. 
Using these 
results we infer the radiative  efficiency in \S \ref{sec:efficiency_and_implications} and  we explore the limits it sets on the emission process.  
We discuss the temporal evolution {of the IC peak} in  \S \ref{sec:evolution} and we conclude in \S \ref{sec:discussion}.

\section{Summary of observations}
\label{sec:observations}

GRB 190114C was a bright burst that was detected by  FERMI/GBM \citep{Hamburg2019}, FERMI/LAT \citep{Kocevski2019} Swift 
\citep{Krimm2019} and Konus-Wind \citep{Frederiks2019}. 
The MAGIC Cherenkov telescope detected the first ever 
sub-TeV  GRB emission from this burst \citep{Mirzoyan2019}. 
We summarize below the main GCN results that are relevant for our work. Note that as these results are preliminary our conclusions may have to be revised  if some of those observational results change significantly during the refined analysis of the data. 

{\bf Redshift:} 
The associated  optical transient  revealed a redshift of  $0.4245$  \citep{Selsing2019,Castro-Tirado2019}.

{\bf Prompt emission:} The prompt emission of GRB 190114C consists of initial hard-spectrum multi-peaked pulse with duration $\simeq 6$~s and a weaker and softer pulse starting $\simeq 16$~s after the trigger 
\citep{Frederiks2019}.
The peak luminosity was $L_{peak} \simeq 1.67 \times 10^{53}$~erg/s \citep{Frederiks2019}.

{\bf Energy release:} According to Fermi/GBM data 
\citep{Hamburg2019},
the prompt energy radiated by GRB 190114C is 
$E^\mathrm{iso}_\mathrm{rad} = 3 \times 10^{53}$~erg (isotropic equivalent). Konus-Wind team reports \citep{Frederiks2019}
somewhat smaller value
$E^\mathrm{iso}_\mathrm{rad} = 2.4 \times 10^{53}$~erg.

{\bf Duration:} The reported duration of GRB 190114C is $T_{90} \simeq 116$~s in $50 - 300$~keV energy range, according to Fermi/GBM team 
\citep{Hamburg2019}, 
and $T_{90} \simeq 361$~s in $15-350$~keV energy range, according to Swift-BAT team \citep{Krimm2019}.
The larger number was also claimed to be a possible underestimate. {The large disagreement between the two estimates of $T_{90}$ may indicate that the burst's emission at $t \gtrsim 100$~s is dominated by slowly decaying afterglow component and that afterglow's contribution to the overall radiated energy is well above 10 per cent.}

{\bf The extended emission:}
As a reference point, we take the moment 70 seconds after the trigger. By this time the prompt emission of GRB 190114C has faded away and Swift-XRT and MAGIC observations started at around this time.
Judging from the Konus-Wind lightcurve\footnote{ \textcolor{blue}{http://www.ioffe.ru/LEA/GRBs/GRB190114\_T75422/}}, the burst's luminosity at this moment was approximately $3 \times 10^{-3}$ of the peak value, that corresponds to  $L_{70sec} \simeq 5 \times 10^{50}$~erg/s. In the available plots, the Swift-BAT lightcurve \footnote{\textcolor{blue}{http://gcn.gsfc.nasa.gov/notices\_s/883832/BA/}} appears saturated at 70 seconds after the trigger, but extrapolation from $150 \div 300$~s time interval gives a comparable number, $L_{70sec}/L_{peak} \sim 4 \times 10^{-3}$.
The energy radiated at later time, $t > 70$~s, can be estimated as $E_{afterglow} \sim L_{70sec} \times 70\, \mathrm{sec} \simeq 3.5 \times 10^{52}$~erg, that is $\simeq 0.15 E^\mathrm{iso}_\mathrm{rad}$. This is likely an underestimate, especially if the lightcurve's decay law is not much steeper than $\propto t^{-1}$. The average flux of GRB 190114C in Swift-XRT energy range in the time interval $70 \div 100$~s after the trigger was $F_X \simeq 1.3 \times 10^{-7}$~erg/s/cm$^2$ (corresponds to fluence $\simeq 3.8 \times 10^{-6}$~erg/cm$^2$) \footnote{\textcolor{blue}{http://www.swift.ac.uk/xrt\_spectra/00883832/}}. Comparing this flux to the fluence reported by Konus-Wind team ($4.83\times 10^{-4}$~erg/cm$^2$,  
\citep{Frederiks2019}
and their estimate for $E^\mathrm{iso}_\mathrm{rad}$, we estimate the  average X-ray luminosity in the time interval $70 \div 100$~s after the trigger {(in the  observer's frame)} as $L^\mathrm{iso}_X \simeq 9 \times 10^{49}$~erg/s {(isotropic equivalent; in the progenitor's frame)}. 

{\bf GeV  observations:}  GRB 190114C was in the Fermi-LAT field of view for 150 seconds since the trigger \citep{Kocevski2019}.
The estimated energy flux above 100 MeV during this period is $\simeq 2 \times 10^{-6}$~erg/s/cm$^2$, that constitutes a fair fraction of the flux at smaller photon energies { \citep[see][for a discussion of GeV emission]{Ravasio19}}.  The highest observed photon energy is 22.9~GeV. This event was observed 15 seconds after the trigger and most likely should be attributed to the prompt emission. In this paper we do not discuss the origin of the observed GeV emission  and we focus on the two dominant afterglow components, the X-rays and the sub-TeV emission.  

{\bf TeV  observations:} The MAGIC Cherenkov telescope detected sub-TeV gamma-ray emission from GRB 190114C \citep{Mirzoyan2019}.
The observation started 50 seconds after the trigger and resulted in detection of point-like source with the significance $>20 \sigma$ in the first 20 minutes. The source was reported to fade quickly. Due to poor observational conditions (large zenith angle $\simeq 60^{\circ}$ and the presence of partial moon) the energy threshold was $\simeq 300$~GeV. For MAGIC sensitivity, we estimate that $20 \sigma$ detection in 20 minutes corresponds to the fluence $\simeq 3 \times 10^{-8}$ erg/cm$^2$. Given the redshift of GRB 190114C, it is beyond the gamma-ray horizon even at the threshold energy (about 1/4 of 300~GeV gamma-rays reach the Earth from the GRB's distance) and therefore only fairly narrow spectral range $300-400$~GeV contributed to the MAGIC fluence. The TeV component is probably $5-10$ times wider (in logarithmic units). After correction for absorption and spectrum's width we estimate that the intrinsic TeV fluence is $\sim 10^{-6}$~erg/cm$^2$.

{
We focus the calculations on a single epoch, the observations at $70$ s after the trigger, corresponding to $50$ s in the local frame. In our numerical estimates below we use the following values, based on this observational summary. 
}

\begin{center}
\begin{tabular}{ l l }
Quantity  & Value in progenitor's frame \\
& (for $z = 0.4245$)\\
\hline
  Time since explosion   & $t = 50$~s \\
 Energy of sub-TeV photons  & ${ E}_\mathrm{_{IC}} = 500$~GeV \\
 \begin{tabular}{l}Prompt radiated energy \\
 (isotropic equivalent) 
 \end{tabular}&  ${ E}^\mathrm{iso}_\mathrm{rad} = 3 \times 10^{53}$~erg \\
 
 \begin{tabular}{l}
 Average isotropic equivalent \\ 
 X-ray luminosity at $t=50$~s
 \end{tabular}   & ${ L}^\mathrm{iso}_X = 9 \times 10^{49}$~erg/s  \\
 Ratio of sub-TeV to X-ray luminosities & $\eta_\mathrm{_{IC}} = 0.25$
\end{tabular}
\end{center}
In the following, unless stated otherwise all quantities are measured in the source frame. All energies and luminosities are isotropic equivalent. We express quantities denoted by $\hat \ $ 
in terms of the observed  values in GRB 190114C, e.g.  ${\hat E}_\mathrm{_{IC}} \equiv E_\mathrm{_{IC}}/ 500 ~{\rm GeV}$. 
With  $20 \div 25$ per cent radiative efficiency at the prompt phase, these numbers correspond to
kinetic energy of ejecta at the afterglow phase  $E^\mathrm{iso}_\mathrm{tot} \simeq 10^{54}$~erg (isotropic equivalent).
The bolometric luminosity of GRB 190114C is larger than the X-ray luminosity, which we infer from the Swift-XRT X-ray data. It includes contribution from soft gamma-ray, MeV, GeV and TeV spectral domains, which is comparable to that of X-ray domain. Somewhat arbitrarily we estimate the bolometric luminosity as {$L^\mathrm{iso}_\mathrm{bol} \equiv \eta_\mathrm{bol} { L}^\mathrm{iso}_X \simeq 2 \times 10^{50}$~erg/s}, and we will use {the correction factor $\eta_\mathrm{bol} \simeq 2$} as a parameter.

\section{The Model}
\label{sec:model} 

The late observations of the sub-TeV component suggests that it arose from the afterglow. We consider, therefore,  an external shock model. Given the scarcity of currently available data a simple one zone model is {acceptable}. The shock is then characterized by  its Lorentz factor, $\Gamma$, {radius, $R$, which can be expressed in terms of $\Gamma$ and the time since the explosion, $t$,} and the surrounding matter density, $\rho$. For simplicity we 
consider a single energy electron population, characterized by the electron's Lorentz factor $\gamma_e$, {but we stress that the results are valid even for  more general electron distributions}.
As justified later in this section we focus here on IC  emission as the source of the sub-TeV emission. 

It is common \citep[see e.g.][]{Piran99} to characterize the condition within the emitting region, the  downstream, using the local equipartition parameters $\epsilon_e$ and $\epsilon_B$ that relate the electron's energy density and the magnetic energy density to the total {downstream energy density}, $e$. 
However, one can use other parameters to characterize the conditions. In particular those parameters can be interchanged with the Compton $y$ parameter, with $ t_\mathrm{cool}/t_\mathrm{dyn}$ the ratio between the {electrons' radiative} cooling time $t_\mathrm{cool}$, and the {shock's dynamical timescale} {$t_\mathrm{dyn} = R/(\Gamma c)$}, and the overall radiation efficiency $\epsilon_r$. {Given} that $y$ {is easy to derive from observations}  for 190114C,  it will be illuminating to use $y$ at times instead of  one of the microphysical parameters to characterize the system. 

There are only two efficient emission mechanisms {for external shock}: synchrotron and {IC}  {from electrons and/or positrons}. Synchrotron is strongly disfavoured as the source of the sub-TeV photons. In the simplest model, the radiating electrons/positrons are accelerated by Fermi mechanism  (diffusive shock acceleration, shear flow acceleration or acceleration in turbulent {electromagnetic} fields) and the rate of energy gain is limited to $\sim e B c$. Equating this rate with the rate of synchrotron losses gives the largest energy a particle can achieve and, therefore, the largest energy of synchrotron photons, the so called burn-off limit
\citep[e.g.][]{Guilbert1983,de-Jager1996,Aharonian2000}: 
{$E \sim m_e  c^2/\alpha$}, where $\alpha$ is the fine structure constant. 
If the sub-TeV photons were Lorentz-boosted synchrotron photons, then the bulk Lorentz factor must be larger than 5000.
For the time of observations this implies unrealistically low density of material around the GRB source.

In principle, there are several ways  to surpass the {burn-off limit} for synchrotron photons. All  use the idea of accelerating electrons in one place, with weaker magnetic field, and then {making} them radiate in {other} regions with a stronger magnetic field \citep[e.g.][]{Kumar2012}.
One  such mechanism  is ultra-fast reconnection with formation of pinch-like structures, where local magnetic field can be much stronger than the average value \citep[see e.g.][]{Kirk2004,Uzdensky2011,Cerutti2012,Kagan2016}.  However, it can occur only in magnetically-dominated environments, which are  unlikely for external shocks.
The converter acceleration mechanism \citep{converter1,converter2} also provides the necessary non-local acceleration. However, in this case the highest-energy photons are less beamed than the low-energy photons \citep{off_axis}, and the sub-TeV synchrotron radiation generated due to converter acceleration should be time-dilated with respect to softer spectral ranges, { unlike} the observations of GRB 190114C.  

We consider therefore, in the following,  Inverse Compton  in the context of SSC, as there is an observed  significant flux of X-ray photons, produced presumably by synchrotron {mechanism} and those are the natural seeds for the IC process. Namely, 
the seed photons likely being the synchrotron radiation of the same electrons.
Within SSC {scenario} the {TeV-}emitting electrons must be {in the} fast cooling {regime} (see \S \ref{sec:fast_cooling}), but even if the X-rays are not produced by the same electrons their large flux would force a fast cooling solution.

\section{The Blast Wave}
\label{sec:auxiliary_relations}

We  use the  theory of an adiabatic blast wave to express 
the physical conditions within the emitting regions in terms of three quantities, the isotropic equivalent {bolometric} luminosity {$L^\mathrm{iso}_\mathrm{bol} \equiv \eta_\mathrm{bol} L^\mathrm{iso}_X$},
the time in the source frame  $t$, and the shock's Lorentz factor $\Gamma$. 
These expressions are well known \citep[e.g.][]{Piran99} (usually in terms of other variables)  and are given here for completeness. 
The necessary expressions are summarized below.

We consider a uniform medium around the progenitor {(ISM for short)} and a stellar wind (wind for short). The density of circum-burst medium, $\rho(R)$, is \begin{equation}\label{eq.rho} 
\rho = \left\{
\begin{array}{ll}
    \displaystyle
     \frac{\dot{M}}{4\pi R^2 v_w} &  \qquad {\rm (wind)} \, , \\[2ex]
    \displaystyle
    \rho_0  & \qquad {\rm (ISM)} \, , 
    \end{array}
    \right.
\end{equation}
where $\rho_0$  is the local density of circum-burst medium,  $\dot{M}$ and $v_w$ are the mass-loss rate and the wind velocity. 

The radius of the blast wave, $R$, and its Lorentz factor $\Gamma$ are related to the observed time $t$ as:
\begin{equation}  \label{R(t)}
    R \simeq  \coeff{4}{8} \Gamma^2 c t  \, .
\end{equation}
Here and in many expressions below, the wind and ISM cases differ only by numerical factors.  We present the results as a single expression preceded by column of two coefficients: the upper one  for a wind and the lower one for an ISM.
{The shock's Lorentz factor}  $\Gamma$ at a given time is expressed in terms of $E^\mathrm{iso}_\mathrm{tot}$: 
\begin{equation} \label{Gamma(t)}
    \Gamma \simeq   \left\{
    \begin{array}{ll}
    \displaystyle
     \left( \frac{E^\mathrm{iso}_\mathrm{tot} v_w}{4\, \dot{M} c^3 t}  \right)^{1/4}
      \, ,  &  \qquad {\rm (wind)} \\[2ex]
    \displaystyle
     \frac{1}{2} \left( \frac{3 E^\mathrm{iso}_\mathrm{tot}}{8\pi \rho_0 c^5 t^3}  \right)^{1/8}
     \, , & \qquad {\rm (ISM)} \, .
    \end{array}
    \right.
\end{equation}

We define the radiative efficiency  $\epsilon_r$  as the ratio of outgoing radiation energy flux to the upstream energy flux:
$\epsilon_r  =
    {\eta_\mathrm{bol} L^\mathrm{iso}_X} / 4\pi R^2  \Gamma^4 \rho c^3 $. 
The magnetic field caries a  fraction $\epsilon_B$ of  the comoving-frame energy density $e = 2 \Gamma^2 \rho c^2$
 and the comoving-frame magnetic field strength is:
\begin{equation} \label{comovingB}
    B = \left( 8\pi \epsilon_B e \right)^{1/2}  \simeq \coeff{1}{1/2}
     \frac{1}{2 \Gamma^3} \left( \frac{\epsilon_B {\eta_\mathrm{bol} L^\mathrm{iso}_X}}{\epsilon_r c^3 t^2} \right)^{1/2} \, .
\end{equation}
Finally, we write the {isotropic equivalent} energy of the {shock} as:
\begin{equation} \label{Etot(L)}
    E^\mathrm{iso}_\mathrm{tot} \simeq \coeff{1}{2/3}    \frac{4 {\eta_\mathrm{bol} L^\mathrm{iso}_X} t}{\epsilon_r} \, .
\end{equation}

\section{Opacity}
\label{sec:TeV_transparency}

The fact that  the sub-TeV photons  have not been absorbed at the source is not trivial. Thus, before turning to the radiation mechanism we  consider the implications of this simple observation. 
The sub-TeV photons  are emitted along with lower-energy X-ray photons. Regardless of origin of the X-ray radiation -- the afterglow emission, the trailing part of prompt emission,  or both -- it could make the source opaque for the sub-TeV photons of energy $E_\mathrm{_{IC}}$ due to two-photon pair production. The main contribution to the opacity comes from photons of energy 
$\sim E_a = 3 \Gamma^2 (m_e c^2)^2/E_\mathrm{_{IC}} \simeq \Gamma^2 \times 1.6\;{\rm eV}$ (in the observer's frame this energy is $E^\mathrm{obs}_a = E_a / (1+z) \simeq \Gamma^2 \times 1.1\;{\rm eV}$).
Let $\eta_a$ be the fraction of the X-ray luminosity  emitted at energies around $E_a$, then the optical depth for absorption of the sub-TeV photons is 
\begin{eqnarray} \label{tau_gg}
    \tau_{\gamma\gamma} &\simeq& \frac{1}{\Gamma^2} \sigma_{\gamma\gamma} n_a R 
    \simeq   \coeff{ 1}{1/2}
  \sigma_{\gamma\gamma} \frac{\eta_a L^\mathrm{iso}_X  E_\mathrm{_{IC}}}{48 \pi \Gamma^6 t c^2 (m_e c^2)^2}    \nonumber \\
 & \simeq   &
    \coeff{ 1.6 }{ 0.8 }
  \frac{\eta_a \hat L^\mathrm{iso}_{X}  \hat E_{\mathrm{_{IC}}}}{\Gamma_2^6 \hat t}
\end{eqnarray}
where $\sigma_{\gamma\gamma} \simeq 0.15\, \sigma_{T}$ is the value of two-photon pair production cross-section near its peak, calculated assuming isotropic distribution of target photons. Recall that $\hat X $ denotes  the value corresponding to the one observed in GRB 190114C (see \S \ref{sec:observations}) and here and elsewhere {$\Gamma_2 \equiv \Gamma/100$.  
Note that this result was obtained here using the observed parameters. However it is more general see  \S \ref{sec:fast_cooling}. 

Using the explicit dependence of the shock's Lorentz factor on observer's time (Eq.~\ref{Gamma(t)}),  we rewrite Eq.~(\ref{tau_gg}) as
\begin{equation} \label{tau_gg(t)}
    \tau_{\gamma\gamma} 
    \simeq   \left\{
    \begin{array}{ll}
    \displaystyle
    \sigma_{\gamma\gamma} \frac{\eta_a  E_\mathrm{_{IC}} c}{6\pi 
    (m_e c^2)^2} \left( \frac{v_w}{c} \right)^{-3/2} \frac{\left( L^\mathrm{iso}_X t \right) \dot{M}^{3/2}}{\left( E^\mathrm{iso}_\mathrm{tot}\right)^{3/2} t^{1/2}} 
    \, ,  &  \qquad {\rm (wind)} \\[3ex]
    \displaystyle
    \frac{8}{9} \left( \frac{6}{\pi}  \right)^{1/4} \sigma_{\gamma\gamma}  \,
    \frac{\eta_a  E_\mathrm{_{IC}} c^{7/4}}{(m_e c^2)^2} \,
    \frac{\left( L^\mathrm{iso}_X t \right) \rho_0^{3/4}}{\left( E^\mathrm{iso}_\mathrm{tot}\right)^{3/4}} \, t^{1/4}
    \, , & \qquad {\rm (ISM)} \, .
    \end{array}
    \right.
\end{equation}
Clearly a source capable of emitting sub-TeV radiation must have  $\tau_{\gamma\gamma} \lesssim 1$. 
This  implies that there is  bias against observing sub-TeV and more energetic emission in dense circum-burst environments both for the wind model, {where} $\tau_{\gamma\gamma} \propto \dot{M}^{3/2}$, and for the ISM model, {where} $\tau_{\gamma\gamma} \propto \rho_0^{3/4}$. For the wind model there is an observational bias against weak bursts, $\tau_{\gamma\gamma} \propto \left( E^\mathrm{iso}_\mathrm{tot}\right)^{-1/2}$, whereas for the ISM model there is a feeble bias in favor of weak bursts, $\tau_{\gamma\gamma} \propto \left( E^\mathrm{iso}_\mathrm{tot}\right)^{1/4}$. The two models differ also in the time dependence of the two-photon absorption optical depth: it slowly decreases with time for the wind model, $\tau_{\gamma\gamma} \propto t^{-1/2}$, and -- even slower -- increases with time for the ISM model, $\tau_{\gamma\gamma} \propto t^{1/4}$. 
Note that when estimating the time dependence we have approximated the shock luminosity  decrease with time as $1/t$ and we have ignored the dependence of  $\eta_a$ on time, $E^\mathrm{iso}_\mathrm{tot}$ and $\dot{M}$ (or $\rho_0$). We don't expect those factors to be { significant} enough to change qualitatively our results. 

The requirement $\tau_{\gamma\gamma} \lesssim 1$ sets a limit on the Lorentz factor:  
\begin{equation} \label{transparency}
    \Gamma \gtrsim  
    \coeff{108}{96}
  \left( \frac{\eta_a \hat L^\mathrm{iso}_{X}  \hat E_{\mathrm{_{IC}}}}{\hat t} \right)^{1/6} \, ,
\end{equation}
Note that $\eta_a$ is a function of $\Gamma$ but because of the weak dependence of $\Gamma$ on all other parameters this can be ignored. For {$\Gamma \simeq 100$} the energy of the absorbing photons is  {$E^\mathrm{obs}_a \simeq 11$~keV} and $\eta_a$ is not much below unity, therefore the transparency condition (\ref{transparency}) is {$\Gamma \gtrsim 100$}. 
{Because of the strong dependence of the opacity on $\Gamma$, the latter conclusion will not change significantly if the source is moderately opaque ($\tau_{\gamma\gamma} \simeq 1 \div 2$), as suggested by our analysis of radiative efficiency in \S \ref{sec:efficiency_and_implications}. Instead, this would make the estimate more certain: $\Gamma \simeq 90 \div 100$ for a moderately opaque source.  
}

{We derived Eq.~(\ref{tau_gg}) assuming} that the low energy (X-ray) photons,  {that absorb}  the high energy (sub-TeV) photons, 
are emitted by the same source. If the low energy photons are prompt radiation that is emitted from a smaller radii then they propagate in small angles relative to the {shock normal}  and this accordingly reduces the interaction rate. However, given the very weak dependence in Eq. \ref{transparency} on the X-ray luminosity (1/6 power) this limit will be more or less valid even if only a small fraction of the X-ray photons is {produced}  by the electrons emitting the high energy radiation. 

The lower limit on the shock's Lorentz factor, set by the transparency condition (Eq.~\ref{transparency}), in combination with shock deceleration law (Eq.~\ref{Gamma(t)}) yields an upper limit on the external density. This corresponds  
to an upper limit on the mass loss rate for the wind case
\begin{eqnarray}
    \dot{M} <  \dot{M}_\mathrm{upp} &=&  \frac{E^\mathrm{iso}_\mathrm{tot} v_w}{c^3 t}  
    \left( \frac{6\pi\,  t c^2 (m_e c^2)^2}{\sigma_{\gamma\gamma} \eta_a L^\mathrm{iso}_X  E_\mathrm{_{IC}}} \right)^{2/3}   \\
    &\simeq& 7 \times 10^{-6} \;
     \frac{E^\mathrm{iso}_\mathrm{tot,54}\, {v_{w,8.5}}}
     {\hat t^{1/3}  \left( \eta_a \hat L^\mathrm{iso}_X \hat E_\mathrm{_{IC}} \right)^{2/3} }
     \;\;  M_{\odot}/\mathrm{yr}\, ,\nonumber
\end{eqnarray}
{where $v_{w,8.5} = v_w / 10^{8.5}$~cm/s,} 
and  to an upper limit on the density of circum-burst medium {for the ISM case}
\begin{eqnarray}
    \rho_0 < \rho_\mathrm{0,upp}& = &\frac{3 E^\mathrm{iso}_\mathrm{tot}}{8\pi c^5 t^3} 
    \left( \frac{3\pi t c^2 (m_e c^2)^2}{2 \sigma_{\gamma\gamma} \eta_a L^\mathrm{iso}_X  E_\mathrm{_{IC}}}  \right)^{4/3}  \\
    &\simeq& 13\; 
    \frac{E^\mathrm{iso}_\mathrm{tot,54}}{\hat t^{5/3} \left( \eta_a \hat L^\mathrm{iso}_X  \hat E_\mathrm{_{IC}}  \right)^{4/3}  } 
    \;\; m_p /\mathrm{cm}^{3} \, .\nonumber
\end{eqnarray}
{In both equations $E^\mathrm{iso}_\mathrm{tot,54} = E^\mathrm{iso}_\mathrm{tot}/10^{54}$~erg.}  
Given these values, which are within the range that is typically expected in both cases, and the weak dependence of $\Gamma$ on the external density (see Eq. \ref{Gamma(t)})  there is no much freedom in the value of $\Gamma$.  Namely $\Gamma$ cannot be much larger than the opacity limit given in Eq. \ref{transparency}.

\section{ Comptonization Regimes}
\label{sec:contra-KN_argument}

The IC mechanism  comes in two varieties: either it operates in Thomson regime, where the energy of the electrons/positrons greatly exceeds the energy of the upscattered photons, or in KN regime, where the energy of the upscattered photons approximately equals the energy of the electrons/positrons. {The observation of GRB 190114C at sub-TeV energy allows us to discriminate between these two options.
} 
Let $E_\mathrm{sy}$ be the photon energy at the synchrotron peak of the SED and $\gamma_e$ the (comoving-frame) Lorentz factor of electrons,  that emits synchrotron photons with this energy. 
 {We define $\gamma_\mathrm{cr}$ as} the critical electron Lorentz factor that satisfies the relation 
\begin{equation} \label{critical_gamma}
m_{e} c^2 = \gamma_\mathrm{cr} E_\mathrm{sy} = \gamma_\mathrm{cr}^3 \hbar \omega_{B} 
\qquad \Rightarrow \qquad
\gamma_\mathrm{cr} = \left( \frac{B_\mathrm{cr}}{B} \right)^{1/3} ,
\end{equation}
where $B_\mathrm{cr} \simeq 4.5 \times 10^{13}$~G is the Schwinger field strength. Electrons with Lorentz factor { $\gamma_e < \gamma_{\rm cr}$} Comptonize their own synchrotron radiation in the Thomson regime, and for $\gamma_e \gtrsim \gamma_{\rm cr}$ Comptonization proceeds in the KN regime.
The largest energy of the IC photons, which can be produced in Thomson regime, is
\begin{equation} \label{crit_Eic}
    E_\mathrm{_{IC}}^\mathrm{cr}  = \Gamma \gamma_\mathrm{cr} m_e c^2 
    = \Gamma \left( \frac{B_\mathrm{cr}}{B} \right)^{1/3} m_e c^2 \, .
\end{equation}
If this energy is larger than the energy of the observed IC photons then   Comptonization is in the Thomson regime, otherwise  it is  in the KN regime.

Substituting the magnetic field strength 
{expected in the external shock (Eq.~\ref{comovingB}) into Eq.~\ref{crit_Eic} }we find that
\begin{eqnarray}
    \label{crit_Eic_evaluated}
    E_\mathrm{_{IC}}^\mathrm{cr} &\simeq&    \coeff{2^{1/3}}{4^{1/3}}
    \Gamma^2 \left( B_\mathrm{cr}^2 \frac{\epsilon_r\, t^2 c^3}{\epsilon_B \eta_\mathrm{bol} L^\mathrm{iso}_X} \right)^{1/6} m_e c^2 \\
   & \simeq& \coeff{0.7}{0.9} \; 
    \Gamma_2^2 \left(  \frac{\epsilon_r}{\epsilon_B {\eta_\mathrm{bol}}} \right)^{1/6} \frac{\hat t^{1/3}}{(\hat L^\mathrm{iso}_X)^{1/6}} \;\; \mathrm{TeV}  \nonumber
\end{eqnarray} 
Due to very weak dependence on the ratio of unknown factors $\epsilon_r$ and $\epsilon_B$ (this ratio is probably {not far from} unity, as suggested by {the comparable} luminosities in synchrotron and IC radiation), equation~\ref{crit_Eic_evaluated} serves as a limit on the Lorentz factor of the external shock that separates the two Comptonization regimes, $E_\mathrm{_{IC}} < E_\mathrm{_{IC}}^\mathrm{cr}$ and $E_\mathrm{_{IC}} > E_\mathrm{_{IC}}^\mathrm{cr}$. If the shock's Lorentz factor is larger than 
\begin{eqnarray} \label{GammaKN}
    \Gamma_\mathrm{_{KN}} &\simeq&    \coeff{2^{-1/6}}{2^{-1/3}} 
    \left( \frac{E_\mathrm{_{IC}}}{m_e c^2} \right)^{1/2} 
    \left( \frac{\epsilon_B {\eta_\mathrm{bol}} L^\mathrm{iso}_X}{\epsilon_r\, B_\mathrm{cr}^2 t^2 c^3} \right)^{1/12}   \\
    &\simeq & \coeff{85}{76} \left( \frac{\epsilon_B {\eta_\mathrm{bol}}}{\epsilon_r} \right)^{1/12}
    \frac{\hat E_\mathrm{_{IC}}^{1/2} (\hat L^\mathrm{iso}_X)^{1/12}}{\hat t^{1/6}} \, \nonumber
\end{eqnarray}
then the {IC} radiation is produced in the Thomson regime. 

{For } GRB 190114C the transparency condition (\ref{transparency}) { implies} that $\Gamma > \Gamma_\mathrm{_{KN}}$ and hence the observed sub-TeV radiation is produced in the Thomson regime, but not very far from the KN limit. 
The fact that $\Gamma_\mathrm{_{KN}}$ is close to the limit set by the transparency condition (\ref{transparency}) is a mere coincidence for the generic afterglow model that employs the classical equipartion parameters \citep{Sari1998}, i.e., 
 the peak of synchrotron spectrum is by chance close to the energy of photons, which contribute most to the opacity. On the contrary, it is  a  basic prediction for pair-balance model \citep{Derishev2016}.

\section{The Lorentz factor of the radiating electrons}
\label{sec:electron_Lorentz_factor}

One may calculate the comoving-frame Lorentz factor, $\gamma_e$,  of the electrons that produce the IC photons 
in two ways: assuming that Comptonization proceeds in KN regime, that gives 
\begin{equation} \label{min_gamma_e}
     E_\mathrm{_{IC}} \simeq \Gamma \gamma_{e,_\mathrm{KN}} m_e c^2 \qquad \Rightarrow \qquad
    \gamma_{e,_\mathrm{KN}} \simeq \frac{E_\mathrm{_{IC}}}{\Gamma m_e c^2}\, ,
\end{equation}
or assuming  that electrons are producing IC photons by upscattering of their own synchrotron photons in  the Thomson regime\footnote{Strictly speaking, {the statement that the largest contribution to the seed photons in Thomson regime is due to self-produced synchrotron photons} is true if the synchrotron SED around the frequency $\gamma_e^2 \omega_B$ has a convex shape in log-log plot.}, that gives
\begin{equation}
    E_\mathrm{_{IC}} \simeq \Gamma \gamma_{e,\mathrm{Th}}^4\, \frac{B}{B_\mathrm{cr}}\, m_e c^2
    \qquad \Rightarrow \qquad 
    \gamma_{e,\mathrm{Th}}  
    \simeq \left( \frac{E_\mathrm{_{IC}}}{\Gamma m_e c^2} \frac{B_\mathrm{cr}}{B} \right)^{1/4} \, .
\end{equation}
From the last equation, substituting the magnetic field strength $B$ from Eq.~(\ref{comovingB}), we obtain
\begin{eqnarray} \label{max_gamma_e}
    \gamma_{e,\mathrm{Th}}  
    \simeq  \coeff{2^{1/4}}{2^{1/2}} 
    \left(\Gamma^2 B_\mathrm{cr} \frac{E_\mathrm{_{IC}}}{m_e c^2}  \right)^{1/4}
    \left( \frac{\epsilon_r c^3 t^2}{\epsilon_B {\eta_\mathrm{bol}} L^\mathrm{iso}_X} \right)^{1/8}
     \, . 
\end{eqnarray}
Comparison of the above equation with Eq.~(\ref{GammaKN}) reveals a simple relation
\begin{equation}  
\label{max_gamma_e(GammaKN)}
    \gamma_{e,\mathrm{Th}} 
    \simeq  \left( \frac{\Gamma}{\Gamma_\mathrm{_{KN}}} \right)^{3/2} \gamma_{e,_\mathrm{KN}} \, .
\end{equation}

The actual value of the electron's Lorentz factor is
\begin{equation}  
    \gamma_e = \max{\left[ \gamma_{e,\mathrm{Th}},\gamma_{e,_\mathrm{KN}} \right]}\, .
\end{equation}
By making this choice, one immediately knows the regime of Comptonization.

Substituting the lower limit (Eq.~\ref{transparency}) 
on the shock's Lorentz factor $\Gamma$ into Eq.~(\ref{max_gamma_e(GammaKN)}) 
and recalling that $\Gamma$ cannot be much larger,
 we find  that {$\gamma_{e,\mathrm{Th}} \simeq   (1.5 \div 2) \; \gamma_{e,_\mathrm{KN}}$}.  This means, that the observed sub-TeV radiation was produced in the Thomson regime (though rather close to KN regime) and hence 
\begin{equation} \label{gamma_e}
    \gamma_e \simeq \gamma_{e,\mathrm{Th}} 
    \simeq \coeff{1.2 }{1.5 } \times 10^4
    \left( \frac{\epsilon_r}{\epsilon_B {\eta_\mathrm{bol}}} \right)^{1/8} \Gamma_2^{1/2}
    \frac{\left( \hat E_\mathrm{_{IC}} \hat t \right)^{1/4}}{\left( \hat L^\mathrm{iso}_X \right)^{1/8}}     \, .
\end{equation}
Given the weak (${\gamma_{e,\mathrm{Th}}} \propto \Gamma^{1/2}$) dependence on the shock's Lorentz factor  and  the fairly narrow allowed range of $\Gamma$,   this expression  provides   a rather good estimate for $\gamma_e$.

\section{The cooling rate}
\label{sec:fast_cooling}

The cooling parameter is the ratio of the radiative cooling time, $t_{\rm cool}$, to the shock dynamical timescale, $t_{\rm dyn} \simeq R/(\Gamma c)$. Using  the magnetic field strength (Eq.~\ref{comovingB}) and the shock radius (Eq.~\ref{R(t)}) we have: 
\begin{eqnarray} \label{eta_e}
     \frac{t_{\rm cool}}{t_{\rm dyn}}
    &= &\frac{6\pi\, m_e c}{\gamma_e \sigma_T  B^2 (1+ \eta_\mathrm{_{IC}})} 
    \frac{\Gamma c}{R} \\
    &\simeq&   \coeff{6}{12}
    \frac{\pi\, \Gamma^5 m_e c}{\gamma_e \sigma_T 
    (1+ \eta_\mathrm{_{IC}})} \left(  \frac{\epsilon_r c^3 t}{\epsilon_B {\eta_\mathrm{bol} L^\mathrm{iso}_X}} \right) 
     \, . \nonumber
\end{eqnarray}
The slowest cooling 
corresponds to the {smaller}   electron's Lorentz factor $\gamma_e=\gamma_{e,_\mathrm{KN}}$ (see Eq.~\ref{min_gamma_e}):
\begin{equation} \label{max_eta_e}
    \bigg(\frac{t_{\rm cool}}{t_{\rm dyn}}\bigg)_\mathrm{max} \simeq   \coeff{6}{12}
    \frac{\pi\, \Gamma^6 \left( m_e c^2 \right)^2}{\sigma_T (1+ \eta_\mathrm{_{IC}}) E_\mathrm{_{IC}} 
    } \left(  \frac{\epsilon_r c^2 t}{\epsilon_B {\eta_\mathrm{bol} L^\mathrm{iso}_X}} \right)  \, .
\end{equation}

Using Eqs.~(\ref{Gamma(t)}) and ~(\ref{Etot(L)}) {to substitute the shock's Lorentz factor $\Gamma$  and  kinetic energy $E^\mathrm{iso}_\mathrm{tot}$,}  we find the fast cooling condition $\bigg(t_{\rm cool}/t_{\rm dyn}\bigg)_\mathrm{max} < 1$ for a wind:}
\begin{eqnarray} \label{fast_cooling_wind}
   & \dot{M}&  > \frac{v_w}{c}
    \left( \frac{6\pi\, \left( m_e c^2 \right)^2  t}{\sigma_T c (1+ \eta_\mathrm{_{IC}}) E_\mathrm{_{IC}}}   \right)^{2/3}
    \left( \frac{{\eta_\mathrm{bol} L^\mathrm{iso}_X}}{\epsilon_B^2 \epsilon_r}  \right)^{1/3} \\ \nonumber
     &\simeq &9 \times 10^{-9} \;
    v_{w,8.5}
    \left( \frac{\hat t}{(1+ \eta_\mathrm{_{IC}}) \hat E_\mathrm{_{IC}}}   \right)^{2/3}
    \left( \frac{{\eta_\mathrm{bol}} \hat L^\mathrm{iso}_X}{\epsilon_B^2 \epsilon_r}  \right)^{1/3} 
     ~  M_{\odot}/\mathrm{yr} 
\end{eqnarray}
 and for an ISM:
\begin{eqnarray} \label{fast_cooling_ISM}
    &\rho_0&   > \frac{c^3}{t^{2/3}}
    \left( \frac{3\, m_e^2}{16 \sigma_T (1+ \eta_\mathrm{_{IC}}) E_\mathrm{_{IC}}}  \right)^{4/3} 
    \left( \frac{\pi\, \epsilon_r}{\epsilon_B^4 {\eta_\mathrm{bol} L^\mathrm{iso}_X}}  \right)^{1/3}     \\ \nonumber
    & \simeq 7.5& \times 10^{-4}\;
    \frac{1}{\hat t^{2/3}}
    \left( \frac{1}{(1+ \eta_\mathrm{_{IC}}) \hat E_\mathrm{_{IC}}}  \right)^{4/3} 
    \left( \frac{\epsilon_r}{\epsilon_B^4 {\eta_\mathrm{bol}} \hat L^\mathrm{iso}_X}  \right)^{1/3}   
    \;\; m_p /\mathrm{cm}^{3} \, .
\end{eqnarray}
Given these small values we conclude that fast cooling regime for GRB 190114C at time moment $t=50$~s is assured.

The fastest cooling 
corresponds to the {larger}  electron's Lorentz factor $\gamma_e=\gamma_{e,\mathrm{Th}}$ (see Eq.~\ref{max_gamma_e}). {We then multiply Eq.~(\ref{max_eta_e}) by $\gamma_{e,_\mathrm{KN}}/\gamma_{e,\mathrm{Th}}$ to obtain $\bigg(t_{\rm cool}/t_{\rm dyn}\bigg)_\mathrm{min}$
and substitute {$\Gamma$} from the equation for the  {two-photon} optical depth (\ref{tau_gg})
to obtain:} 
\begin{eqnarray} \label{min_eta_e}
    \bigg(\frac{t_{\rm cool}}{t_{\rm dyn}}\bigg)_\mathrm{min} &\simeq&   
     \frac{1}{8 \tau_{\gamma\gamma}}
    \left( \frac{\gamma_{e,_\mathrm{KN}}}{\gamma_{e,\mathrm{Th}}} \right)
    \frac{\sigma_{\gamma\gamma}}{\sigma_T}  \,
     \frac{\eta_a }{ (1+ \eta_\mathrm{_{IC}})  }  \,
     \frac{L^\mathrm{iso}_X}{L^\mathrm{iso}_\mathrm{bol}}  \, 
     \frac{\epsilon_r }{\epsilon_B}     \nonumber \\
     &\simeq&  0.02 \; \frac{1}{\tau_{\gamma\gamma}}
    \left( \frac{\gamma_{e,_\mathrm{KN}}}{\gamma_{e,\mathrm{Th}}} \right)
     {     \frac{\eta_a \epsilon_r}{ (1+ \eta_\mathrm{_{IC}}) \eta_\mathrm{bol} \epsilon_B}  }
 \end{eqnarray}
for both wind and ISM cases.
The similarity of the expression for $\bigg(t_{\rm cool}/t_{\rm dyn}\bigg)_\mathrm{max}$ (Eq.~\ref{max_eta_e}) to expression (\ref{tau_gg}) for the photon absorption optical depth, which eventually leads to the above simple relation, is {not coincidental.  } 
 Both describe electromagnetic interaction of energetic particles (electrons in one case and photons in the other) having the same energy with {the same}  background low-energy photons. The only difference  is in the cross-sections for electron-photon and photon-photon interactions.
Thus, {if Comptonization operates in the KN regime or close to it, then the} IC photons arising from fast  cooling  electrons  should have an optical depth to pair creation with  the low energy seed photons that is {larger than}  $y$ \citep[see e.g][]{cooling_vs_transparensy1,cooling_vs_transparensy2}.  

In the above discussion the cooling rate was considered within the {SSC} model. However, it is important to note that the large flux of low energy X-rays {ensures} that the IC emitting electrons would be in fast cooling regardless of the origin of the X-ray photons{ and} a slow cooling IC {regime} is ruled out. 

\section{The Radiation Efficiency and its Implications}
\label{sec:efficiency_and_implications}
{Assuming turbulent, i.e., isotropic on large scales, magnetic field in the downstream the intensity of the synchrotron radiation at the front of a plane-parallel emitting region at  an angle $\theta$ to its normal is:
\begin{equation} \label{SyIntensity} 
I_\mathrm{sy}(\theta) 
\simeq \frac{1}{\cos(\theta)} 
 \frac{\sigma_T \gamma_e^2 n_e e_B c l_\mathrm{em} }{3\pi} = \frac{I_\mathrm{sy}(0)}{\cos(\theta)} =   y \frac{e_B c}{4\pi \cos(\theta)} \, , \nonumber
\end{equation}
Here $l_\mathrm{em}$ is the thickness of the emitting region measured in the shock  comoving frame: 
\begin{equation} \label{eq:lem}
l_\mathrm{em} \simeq  \begin{cases}
c t_\mathrm{cool} < R/\Gamma  &  \mbox{for } \ \ \ t_\mathrm{cool}<t_\mathrm{dyn}  \ \ \mbox{(fast cooling)}, \\
R/\Gamma & \mbox{for } \ \ \ t_\mathrm{cool}/t_\mathrm{dyn} > 1 \ \ \mbox{(slow cooling)} \ ,
\end{cases}
\end{equation}
and
\begin{equation} \label{def_ComptonY} 
y  \equiv  \frac{4}{3} \sigma_T \gamma_e^2 n_e l_\mathrm{em} 
\end{equation}
is the Compton $y$ parameter.
The energy density of the synchrotron radiation inside the emitting region is
\begin{equation}
e_\mathrm{sy} \simeq \frac{2\pi}{c} \int_0^{\pi/2} I_\mathrm{sy}(\theta) \, \sin\theta \mathrm{d} \theta
\simeq \frac{2\pi \Lambda I_\mathrm{sy}(0)}{c} \simeq \frac{\Lambda}{2} y e_B
 \, .
\end{equation}
The integral in this equation has a logarithmic divergence at $\theta \rightarrow \pi/2$, that is an artifact of the plane geometry approximation. 
However, the  shock  has a finite curvature and hence the integral is finite. We take this into account introducing the geometrical factor 
\begin{equation} \label{geom_factor} 
\Lambda   \simeq   1 + \ln \left( \frac{R}{\Gamma l_\mathrm{em}} \right) \ , 
\end{equation}
that reproduces both asymptotic limits, $\Lambda \simeq 1$ for $l_\mathrm{em} \simeq R/\Gamma$ and $\Lambda \simeq \ln \left( R/\Gamma l_\mathrm{em} \right)$ for $l_\mathrm{em} \ll R/\Gamma$.}

The  synchrotron radiation flux at the shock front is\footnote{
This flux is calculated for a static emission zone. In the case of GRB afterglows, the emitting zone is associated with the downstream plasma, which recedes from
the shock front, and hence fewer photons move in forward direction in the shock plane. On the other hand, most of the photons, which appear to move backwards in the shock frame, actually move in forward direction in the progenitor's frame and overtake the shock at later time when it decelerates.
}
\begin{equation} \label{SyFlux}
F_\mathrm{sy} \simeq 2\pi \int_0^{\pi/2} \cos\theta I_\mathrm{sy}(\theta) \, \sin\theta \mathrm{d} \theta
\simeq 2\pi I_\mathrm{sy}(0) \simeq \frac{y}{2} e_B  c \, .
\end{equation}
Comparing it to the energy flux, associated with the downstream plasma, ${F} = (c /2) e  = (c /2) e_B/\epsilon_B$ for a downstream velocity equal to $c/3$, 
we introduce synchrotron radiative efficiency \citep{Sari1996}:
\begin{equation} 
\label{eq.eff}
\epsilon_\mathrm{sy} \equiv \frac{F_\mathrm{sy}}{{F}} \simeq y \epsilon_B\,  .
\end{equation}
Note that the radiative efficiency can also be expressed in terms of $\epsilon_e$:
\begin{equation} 
\label{eq.effe}
\epsilon_\mathrm{sy} < \epsilon_r =
\epsilon_e \begin{cases}
1 & \mbox{for } \ \ \ t_\mathrm{cool}< t_\mathrm{dyn}  \ \ {\mbox{(fast cooling)}}  , \\
t_\mathrm{dyn}/t_\mathrm{cool}   &    \mbox{for } \ \ \ t_\mathrm{cool}>t_\mathrm{dyn} \ \ {\mbox{(slow cooling)}} .
\end{cases}
\end{equation}
While the  latter expression for the efficiency,  ${\epsilon_r =} \epsilon_e \min(1, t_\mathrm{dyn}/t_\mathrm{cool})$, is more familiar, the  expression  ${\epsilon_\mathrm{sy} =} y \epsilon_B$ is also  useful here as $y$ can be directly estimated from the observable  $\eta_{_\mathrm{IC}}$. 

Calculating the IC radiation flux  at the shock front in the same way as for synchrotron radiation,
 where the magnetic field energy density is replaced by the energy density of synchrotron radiation  {with} an additional factor $\kappa_{_\mathrm{KN}} \leq 1$ that accounts for the KN effect
 we obtain
\begin{equation} \label{IC_Flux}
F_{_\mathrm{IC}} 
\simeq  \kappa_{_\mathrm{KN}} \frac{y}{2} e_\mathrm{sy}  c 
\simeq  \kappa_{_\mathrm{KN}} \frac{\Lambda}{4}  y^2 e_B  c 
\, . 
\end{equation}

Therefore, the IC radiative efficiency is {\citep{Sari1996}}\footnote{Note the geometrical factor $\Lambda$ that was included here relative to\citep{Sari1996}} : 
\begin{equation} \label{ComptonY_proxy}
\epsilon_{_\mathrm{IC}} \equiv \frac{F_{_\mathrm{IC}}}{{F}} \simeq   \kappa_{_\mathrm{KN}} \frac{\Lambda}{2}  y^2 \epsilon_B 
\, ;
\qquad 
\eta_{_\mathrm{IC}} \equiv \frac{F_{_\mathrm{IC}}}{F_\mathrm{sy}} \simeq  \kappa_{_\mathrm{KN}} \frac{\Lambda}{2}  y
\, .
\end{equation}
Note that within the Thomson regime $ y$ and $\eta_{_\mathrm{IC}}$ are the same up to the {logarithmic} geometrical factor $\Lambda/2$. 
{The overall radiative efficiency is} ${\epsilon_r =} \epsilon_\mathrm{sy} + \epsilon_{_\mathrm{IC}} \leq \epsilon_e$. The equality {$\epsilon_r = \epsilon_e$} holds for the fast cooling regime.

If  $L^\mathrm{iso}_\mathrm{IC}$, inferred from the available observational data, is treated as the intrinsic IC luminosity of the external shock, then the Compton $y$ parameter can be estimated from the ratio of sub-TeV (IC) to X-ray (synchrotron) luminosities (see Eq.~\ref{ComptonY_proxy}). Assuming a geometrical factor $\Lambda \sim 2$ and keeping in mind that  Comptonization proceeds in nearly Thomson regime, we arrive at {$y \sim 0.25$}. To first order in $y$, the external shock efficiency can be estimated as {$\epsilon_r \simeq y \epsilon_B \sim 0.25 \epsilon_B$}. 

Using the above estimates for the radiative efficiency and combining it with the expression for the shock's isotropic equivalent kinetic energy (see Eq.~\ref{Etot(L)}) we find that
\begin{equation} \label{Etot(eps)}
   \epsilon_B     \simeq \coeff{1}{2/3}   
     \frac{4 {\eta_\mathrm{bol} L^\mathrm{iso}_X} t}{\left(1  + \eta_{_\mathrm{IC}} \right) y E^\mathrm{iso}_\mathrm{tot}}     
     \simeq \coeff{0.07}{0.05}  
     \frac{\eta_\mathrm{bol} \hat L^\mathrm{iso}_X \hat t}{\hat \eta_{_\mathrm{IC}} E^\mathrm{iso}_\mathrm{tot,54}}  \;\;  \, ,
\end{equation}
where the last approximate equality is valid for small values of $\eta_{_\mathrm{IC}}$ as in the case of GRB 190114C.
{The above equation suggests}  either a large magnetization or a very large value of $E^\mathrm{iso}_\mathrm{tot}$
and consequently a low efficiency of the prompt phase. 

{Large magnetization} is unexpected as the shock propagates into an unmagnetized medium.  In particular,  PIC simulations 
consistently show $\epsilon_B \sim 10^{-2}$ for a shock propagating into unmagnetized medium \citep[e.g.][]{Magnetization1, Magnetization2}. This value implies  $E^\mathrm{iso}_\mathrm{tot} \sim 10^{55}$~erg, the implied real energy  is uncomfortably large, even after adding typical beaming corrections. 
This also implies a radiative efficiency of only a few percent at the prompt phase. Numerical simulations also suggest that the magnetic field's energy share is several times less than that of the accelerated electrons \citep[e.g.][]{ElectronEnergyShare1, ElectronEnergyShare2}. Again this is  at odds with our estimate for GRB 190114C unless the electrons radiate in the slow cooling regime, which is ruled out  {(see Eqs. (\ref{fast_cooling_wind}) and (\ref{fast_cooling_ISM}) and discussion thereafter)}. 

A possible resolution of this apparent problem  is that the sub-TeV radiation is stronger than what we use as a canonical value and $y$ is larger. This can happen if the sub-TeV radiation is strongly absorbed within the emitting zone or if 
our estimate of the sub-TeV luminosity from the available GCN data was too low   or both. 
A larger intrinsic sub-TeV luminosity (and hence a larger Compton $y$) 
would resolve   both  problems of too large  shock's kinetic energy and  too low {$\epsilon_e$} (relative to {$\epsilon_B$)}. 
While a careful examination of the {observational data} could reveal a better estimate for the ratio  of {sub-TeV} to {X-ray} fluxes it may be much more difficult to assess directly  whether there was some level of {internal}  self-absorption of the sub-TeV photons or not. 

Yet another possibility is  that at {an observer time of $t=70$~s}  the X-ray radiation is still dominated by the prompt emission and the external shock's contribution to the observed X-ray and sub-TeV fluxes is small. This would relax the requirements for the shock's kinetic energy.  But the estimate of Compton $y$ parameter would remain essentially unchanged, hence $\epsilon_B > \epsilon_e$ will still hold.

{An independent}  way of estimating the {parameters of the emitting zone is based on the relation between} radiative efficiency {and} the total energy of the radiating {particles (electrons and/or positrons)}.
Given {their} Lorentz factor $\gamma_e$  and their number per  baryon,  $\xi_e$ (note that here $\xi_e$ can be larger than unity if there is a significant pair loading {as a result of internal absorption of IC photons}) 
one can set an  upper limit to the radiative efficiency: 
\begin{equation}
\label{eq:eqe}
  \epsilon_r \leq \epsilon_e \equiv \xi_e \gamma_e m_e / (\Gamma m_p)  
\end{equation}
that {becomes an equality in the fast cooling case} relevant to GRB 190114C. 

Using this relation with Eqs. \ref{eq.eff} and \ref{ComptonY_proxy} we find that in the fast cooling regime
\begin{equation}
\epsilon_B  
\simeq \frac{\xi_e \gamma_e m_e}{y \left(1+\eta_{_\mathrm{IC}} \right) \Gamma m_p}  
\simeq {0.22} \, \frac{\xi_e \gamma_{e,4}}{{\hat \eta_{_\mathrm{IC}}} \Gamma_2}  \, .
\label{eq:eq}
\end{equation}
For the parameters of GRB 190114C we get $\epsilon_B \simeq 0.2 \xi_e$. 
If all the  electrons {from the circum-burst medium}  are accelerated, then $\xi_e = 0.5$ for a Wolf-Rayet stellar wind case and $\xi_e = 0.87$ for the ISM case. This would imply a rather large $\epsilon_B \simeq 0.1 \div 0.2$.
If on the other hand $\epsilon_B$ is  small, as implied by the PIC simulations, {then} the fraction of accelerated of electrons, $\xi_e$,  should be small as well. This is also observed in PIC simulations \citep{Magnetization1}.

Combined with Eq.~\ref{Etot(L)}, 
Eq. \ref{eq:eq} gives a lower limit on the kinetic energy of the  shock.
Comparing this value with the estimated isotropic equivalent prompt $\gamma$-ray energy ($3 \times 10^{53}$~erg from section \ref{sec:observations}) 
the ISM scenario is consistent with a prompt radiation efficiency up to $\simeq 40$ per cent. A smaller prompt efficiency would imply that only a fraction of the available electrons is  accelerated by the external shock. 
In the wind scenario the prompt efficiency is limited to $\lesssim 20$ per cent, unless there are additional electron-positron pairs that are produced within the external shock and are accelerated along with electrons from the wind. 
If one assumes that only a fraction of the observed X-ray luminosity is due to external shock (as discussed earlier in this section), then the shock's contribution to  IC luminosity and hence the requirements on the shock's kinetic energy would be proportionally smaller.

{ The parameters, which we determined for the early afterglow phase of GRB 190114C, are consistent with the {generic afterglow} 
model, albeit with a larger than expected  $\epsilon_B$ value and with and $\epsilon_B > \epsilon_e$ {(if one assumes that the observed ratio of IC to synchrotron luminosity is intrinsic to the source)}.   At the same time they fit  well into more specific predictions of the pair-balance model. 
Two key predictions of this model are: (i)  The IC peak is produced at the border between KN and Thomson regimes;  
(ii) $\eta_{_\mathrm{IC}} = \mbox{a few}$. Both are satisfied here (see Eqs. \ref{GammaKN} and \ref{max_gamma_e(GammaKN)} and discussions there for the first condition). As noted earlier the unexpectedly large inferred magnetization together with the inequality $\epsilon_B > \epsilon_e$ suggest that there is moderate internal absorption of sub-TeV photons. If so this will increase the estimate of the intrinsic Compton $y$ parameter to $\simeq$ a few, getting it closer to the range predicted by the pair balance model. }

\section{Temporal evolution of IC peak position: generic vs. pair-balance models}
\label{sec:evolution}

Predictions of the  temporal evolution of the IC peak are drastically different in these two {models}, making it possible to distinguish between the two  if observations at later time become available. 
In the simplest scenario for both models, the microphysical parameters $\epsilon_B$ and $\epsilon_e$  remain constant. In the case of fast cooling and Comptonization in Thomson regime, this implies that $\eta_\mathrm{_{IC}}$ does not change with time. But the two models differ in the predicted evolution of the IC peak energy, $E_\mathrm{_{IC},p} \propto \Gamma \gamma_{e,\mathrm{p}}^4 B$ (in Thomson regime).

In the generic model {$\gamma_{e,\mathrm{p}}$}  is proportional to $\Gamma$ (see Eq. \ref{eq:eqe}).
Therefore, 
\begin{equation} \label{generic_IC_peak}
    E_\mathrm{_{IC},p} \propto \Gamma^5 B \propto \Gamma^2 t^{-3/2}
    \propto   \left\{
    \begin{array}{ll}
    \displaystyle
    t^{-2}
    \, ,  &  \qquad {\rm (wind)} \\
    \displaystyle
    t^{-9/4}
    \, , & \qquad {\rm (ISM)} \, ,
    \end{array}
    \right.
\end{equation}
where we substituted the magnetic filed strength $B$ from Eq.~(\ref{comovingB}) and then the shock Lorentz factor $\Gamma$ from Eq.~(\ref{Gamma(t)}).
Both for the wind and for the ISM cases the generic model predicts a fast decrease   of the peak IC energy. For example, if a GRB starts with $E_\mathrm{_{IC},p} \simeq 1$~TeV at 100 seconds after the explosion, then one hour later the IC peak would be located at $\simeq 1$~GeV.

In the pair-balance model the peak Lorentz factor of the radiating electrons is determined by the pair-production condition and the number of energetic electrons (positrons) is regulated in such a way that $\gamma_{e,\mathrm{p}} \simeq \gamma_\mathrm{cr} \propto B^{-1/3}$ (see Eq.~\ref{critical_gamma}). Therefore, 
\begin{equation} \label{pairbalance_IC_peak}
    E_\mathrm{_{IC},p} \propto \Gamma B^{-1/3} \propto \Gamma^2 t^{1/2}
    \propto   \left\{
    \begin{array}{ll}
    \displaystyle
     const
    \, ,  &  \qquad {\rm (wind)} \\
    \displaystyle
    t^{-1/4}
    \, , & \qquad {\rm (ISM)} \, .
    \end{array}
    \right.
\end{equation}
The pair-balance model predicts that the peak IC energy does not change with time in the wind case. In the ISM case, the model predicts weak evolution IC peak towards lower energies.  For the same hypothetical GRB, which starts with $E_\mathrm{_{IC},p} \simeq 1$~TeV at 100 seconds after the explosion, the IC peak would be {above} $\simeq 200$~GeV  even at 10 hours after the  explosion and will  still be  accessible for Cherenkov telescopes,
provided that the flux, that decreases like $t^{-1}$, doesn't fall below the {sensitivity limit}.

\section{Conclusions}
\label{sec:discussion}

MAGIC's {observations} of the sub-TeV emission from GRB 190114C opened a new window on the emission process in GRBs' afterglows. Within the SSC framework  this emission has to be assigned to the IC component. It  is the first time when this component was unequivocally observed. With this information at our disposal we are able to constrain the conditions within the emitting region of a GRB to a better precision than has been ever possible.

Our analysis is based on the preliminary data described in GCNs. {We expect that our results will hold unless these values will be significantly revised in the refined analysis. }  Given the available data we use a single zone model and we {do not} attempt to reproduce the whole spectrum. Instead we focus on the two dominant components,  the sub-TeV radiation and  the lower energy X-rays, that turn out to be the  seed photons for the IC process producing the sub-TeV photons. 
 An   external shock with a bulk Lorentz factor $\Gamma \simeq 100$ and electrons accelerated to $\gamma_e \simeq 10^4$ can explain the observations  with a SSC model in which  the IC process is in the Thomson regime but near the transition to the KN regime. The radiating electrons cool rapidly (fast cooling). {Regardless of the details of the model the strong X-ray emission will essentially lead to fast cooling of the IC emitting electrons. } 

{We find that the transparent (for sub-TeV photons) solution is possible at expense of assuming very large shock magnetization or very large kinetic energy. However the optical depth for internal absorption of sub-TeV photons in any case exceeds the value $\left( \Gamma^2 \epsilon_r \right)^{-1}$, that means the upstream acquires enough momentum from secondary pairs to start moving at relativistic speed even before the shock comes. This forces one to use a modified shock solution, as discussed in \citep{Derishev2016}}.

The {detection of sub-TeV photons implies} that the {source's optical depth with respect to two-photon pair production is at most a few}. Note that we cannot exclude absorption of sub-TeV radiation at a moderate level within the source itself. The  {target} photons {for absorption} are  in the X-rays. The pair annihilation opacity is alleviated by the Lorentz boost, just like in the common  compactness  argument \citep[e.g.][]{BaringHarding,Piran99,Lithwick2001}. 
It turns out that for both wind and ISM a minimal bulk Lorentz factor of order $\Gamma \simeq 100$ at the time of observation is needed to allow escape of sub-TeV photons. As usual in compactness arguments the dependence of this limit on the different parameters is rather low and the limit is very robust. 
For the same reason the limit doesn't vary much if we require an optical depth  of a few instead. 
On the other hand the shock deceleration dynamics implies, for reasonable circum-burst densities, that the bulk Lorentz factor must be  $\Gamma \lesssim 100$ 
at the time of the observations.  { Combined with the opacity limit we find  that  the bulk Lorentz factor of the afterglow {is} $\Gamma \simeq 100$. }
 
The electrons must be energetic enough to produce the sub-TeV photons. With $\Gamma \simeq 100$  this implies  $\gamma_e \gtrsim  10^4$.  {An upper limit  $\gamma_e \lesssim 1.2 \div 1.5 \times  10^4$ derives from condition that the emission process is  SSC in the Thomson regime.}  Once more, the two limits bracket $\gamma_e$ nicely from above and from below. 
 Thus, the IC operates in the Thomson regime but very close to the Thomson/KN boundary.  {The seed photons are X-rays  and they are, indeed, the synchrotron emission produced by $\gamma_e \simeq 10^4 $ electrons. Remarkably the observed flux of the  X-ray photons is compatible with this interpretation. } {Furthermore, our analysis indicates that the observed sub-TeV emission is near the peak of the IC component.}

We obtained the values for $\Gamma$ and $\gamma_e$ in a way, that does not use spectral information (i.e., we were not making spectral fits). Yet we arrived at pretty certain estimates. This was possible because we could constrain the IC mechanism to Thomson regime of operation. In this regime the electrons which are responsible for the peak of synchrotron SED comptonize mostly their own synchrotron radiation. This alleviates the major uncertainty of SSC modelling -- a possibility that the seed photons for the main (in the sense of energetics) part of electron distribution are produced by some lower-energy electrons.

If the ratio between IC and synchrotron luminosities $\eta_\mathrm{_{IC}} \simeq 0.25$, as  we used in our estimates, reflects the intrinsic conditions in the emitting zone, then either the GRB's kinetic energy was in excess of $10^{55}$~erg and its radiative efficiency {was}  below several per cent or the   shock magnetization is large, with $\epsilon_B \sim 0.1$. In either case the energy share of the radiating electrons is $\simeq 4$ times smaller than that of the magnetic field. These findings  concerning the microphysical equipartition parameters  depart from {both theoretical expectations  and  results of PIC simulations \citep[e.g.][]{Magnetization1, Magnetization2}}. However, they can be made consistent with those expectations assuming a moderate (with $\tau_{\gamma\gamma} \simeq 2$) intrinsic absorption of the sub-TeV radiation.

If the sub-TeV radiation from GRB 190114C was indeed partially  self-absorbed, {as  suggested by our analysis,}
then we can speculate  
that other bursts regularly escape detection by Cherenkov telescopes just because the sources are typically self-absorbed in the TeV range.
Indeed, the opacity argument  sets an upper limit on the surrounding matter density. While this limit is not very stringent for an ISM, it is rather low for a wind, $\dot{M} <  6.5 \times 10^{-6} E^\mathrm{iso}_\mathrm{tot,54} {v_{w,8.5}} \; M_{\odot}/\mathrm{yr}$, and the majority of progenitors may fail to pass the self-absorption filter  \citep[see also][]{Vurm2017}.  

Under the assumption of moderate intrinsic absorption of the sub-TeV radiation, the conditions in the emitting zone fit nicely into the predictions of pair-balance model: Comptonization proceeds at the border between Thomson and KN regimes;   internal absorption of IC photons provides secondary pairs for further acceleration and emission; the Compton $y$ parameter is {of order unity}.   The  same conditions are possible for a generic model as well, though there is no special preference for this region in the parameter space. A clear distinction between the generic and the pair-balance models can be made if late-time observations of TeV emission become available: the generic model predicts a rapid decline of peak IC energy with time, whereas the pair-balance model {predicts} that the peak IC energy {stays approximately constant in} time. 

The main uncertainty in the interpretation of our results arises from the uncertainty in the {IC-to-synchrotron luminosity} ratio $\eta_\mathrm{_{IC}}$  that we inferred from the preliminary data to be $\approx 0.25$. This has lead to the conclusion that $\epsilon_B > \epsilon_e$ and to the conclusion that $\epsilon_B$ is rather large compared to expectations. However  the analysis outlined here doesn't depend  on this value.  
Clearly the qualitative conclusions will have to be revised if it turns out that $\eta_\mathrm{_{IC}} > 1 $. However the rest of the analysis concerning the conditions within the emitting regions still holds.

Future observations of GRBs in the sub-TeV range  will provide further insight into the  conditions within GRBs' emitting zones. 
In particular we will be able to explore the range of microphsyical parameters that arise in GRBs afterglow. 
Once  the sub-TeV spectra become available, they may shed more light on whether there is significant internal absorption or {not, that} is critical to some parts of the analysis. 
The observations will  enable us to  distinguish between different acceleration mechanisms and explore  the microphsyics of shock accelerations. Beyond GRBs, these   results will  have impact on a whole suite of other astrophysical phenomena involving relativistic shocks.

\section{Acknowledgements}

This research is supported by the Russian Science Foundation grant No 16-12-10528 (ED), by an advanced ERC grant (TREX) and  by the I-Core center of excellence of the CHE-ISF (TP).

\bibliographystyle{mnras}

\end{document}